\documentclass[fleqn,usenatbib]{mnras}

\usepackage{newtxtext,newtxmath}

\usepackage[T1]{fontenc}
\usepackage{ae,aecompl}

\usepackage{graphicx}	
\usepackage{subcaption}
\usepackage{amsmath}	
\usepackage{bm}
\usepackage[usenames]{color}
\usepackage{animate}
\usepackage{mathrsfs}

\graphicspath{{Figures/}}
\usepackage[normalem]{ulem}
\usepackage{array}
\usepackage{framed}

\title[Multi-plane lensing in wave optics]{Multi-plane lensing in wave optics}

\author[Feldbrugge]{Job Feldbrugge$^{1}$\thanks{E-mail: Job.Feldbrugge@ed.ac.uk}
\\
$^{1}$Higgs Centre for Theoretical Physics, James Clerk Maxwell Building, Edinburgh EH9 3FD, UK
}

\date{Accepted XXX. Received YYY; in original form ZZZ}

\pubyear{2022}

\begin{document}
\label{firstpage}
\pagerange{\pageref{firstpage}--\pageref{lastpage}}
\maketitle

\begin{abstract}
    Wave effects in lensing form a rich phenomenon at the intersection of classical caustic singularities and quantum interference, yet are notoriously difficult to model. A large number of recently observed pulsars and fast radio bursts in radio astronomy and the prospected increase in sensitivity of gravitational wave detectors suggest that wave effects will likely be observed in the near future. The interference fringes are sensitive to physical parameters which cannot be inferred from geometric optics. In particular, for multi-plane lensing, the pattern depends on the redshifts of the lens planes. I present a new method to define and efficiently evaluate multi-plane lensing of coherent electromagnetic waves by plasmas and gravitational lenses in polynomial time. This method will allow the use of radio and gravitational wave sources to probe our universe in novel ways. 
\end{abstract}




\begin{figure*}
    \centering
    \begin{subfigure}[b]{0.48\textwidth}
    \includegraphics[width =\textwidth]{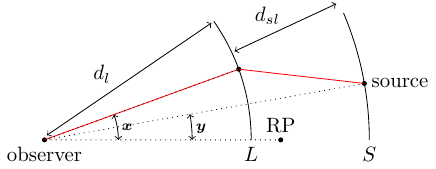}
    \caption{Geometry of interfering paths in a single-plane lens system.}\label{fig:single}
    \end{subfigure}
    \begin{subfigure}[b]{0.48\textwidth}
    \includegraphics[width = \textwidth]{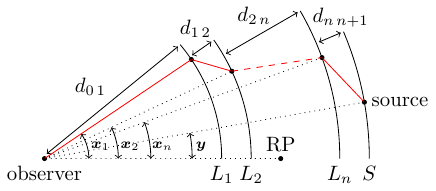}
    \caption{Geometry of interfering paths in a multi-plane lens system.}\label{fig:multi}
    \end{subfigure}
    \label{fig:LensSetup}
    \caption{Single and multi-plane lens systems.}
\end{figure*}

\section{Introduction} \label{sec:intro}
Interference is one of the most universal phenomena in nature. Famous examples are the beats occurring in the superposition of sound waves in acoustics, the color patterns on the surface of a soap bubble, the emergence of fringe patterns in the lensing of electromagnetic radiation in wave optics, and the occurrence of the fringes in Young's double-slit experiment in quantum mechanics. Unfortunately, interference phenomena are often mathematically delicate to define and difficult to evaluate, as their definition frequently involves conditionally convergent multi-dimensional oscillatory integrals\footnote{An integral $\int\! f(x)\mathrm{d}x$ converges conditionally when the integral converges, for some regularization scheme, but the integral over the absolute value of the integrand diverges, \textit{i.e.}, $\int|f(x)|\mathrm{d}x = \infty$. If the latter integral is finite, it is called absolutely convergent. Important theorems such as the dominated convergence theorem and Fubini's theorem do not apply to conditionally convergent integrals, making their definition more subtle.}. 
 Recently, Picard-Lefschetz theory was introduced to theoretical physics in an attempt to unambiguously define and efficiently evaluate oscillatory integrals \citep{Witten:2010}. This theory exploits Cauchy's theorem on the deformation of integration contours in complex analysis and finds the optimal deformation of a general oscillating integral. \cite{Feldbrugge:2019} applied these techniques to the single-plane lens problem in wave optics, constructing its unambiguous definition and an efficient way to evaluate the interference patterns in a general (analytic) one- and two-dimensional single-plane lens.
 
 Lensing is of key importance in astronomy and cosmology, as it allows us to infer properties of systems that cannot be measured in other ways.  For example, gravitational microlensing allows us to detect faint but massive foreground objects, including exoplanets, and to put tight constraints on their contribution to the dark matter \citep{Paczynski:1986, Mao:1991, Gould:1992, Wyrzykowski:2011, Mao:2012, Gaudi:2012}. Plasma lenses have been observed to amplify astronomical signals, such as the ``Black Widow'' pulsar \citep{Main:2018}. The turbulent interstellar medium leads to scintillation effects \citep{Rickett:1977, Rickett:1990, Narayan:1992}.
 
 In these examples, the single-plane lens model provides a good approximation. Yet, radiation can also propagate through multiple lenses before reaching us. When the lenses are widely separated in the radial direction, the single-plane lens model fails and we enter the realm of multi-plane lens systems \citep{Blandford:1986, Schneider:1992, Petters:1995, Petters:2001, Pen:2006, Mao:2012, Yamamoto:2017}. The study of multi-plane lens systems requires the evaluation of high-dimensional oscillatory integrals, to which the techniques proposed by \cite{Feldbrugge:2019} are not well-suited due to scalability. 
 Therefore, in this paper, I propose new methods to efficiently evaluate multi-plane lens systems in wave optics.
 
Why is it important to study systems in wave optics? In astronomy, lensing is often considered in the limit of geometric optics. In this limit, light travels along null geodesics and the image is determined by the geometry of the light rays. This is an excellent approximation in case the wavelength of the radiation is short compared to the characteristic lensing scale, the source is extended, or the radiation is incoherent \citep{Jow:2022}. However, the approximation does not apply universally. For pulsars and fast radio bursts (FRBs), the wave nature of their radiation may be relevant, as these sources are very small and their radio emissions are coherent.  This is also true for gravitational radiation from coalescing binaries \citep{Takahashi:2003, Christian:2018, Meena:2020}.

Moreover, the observation of wave effects in lensing opens a new window for astronomy, as the diffraction pattern of a lens is significantly richer than the corresponding intensity pattern in geometric optics. Unlike the latter, the diffraction pattern depends on the frequency in a way that can be used to constrain the system's physical parameters. For example, the diffraction patterns in the lensing of gamma-ray bursts (GRBs) have been proposed as a probe to constrain possible dark matter candidates \citep{Gould:1992b, Katz:2018}. It was recently demonstrated that wave effects generally increase the cross section for lensing events and allow us to put tight constraints on the mass of a gravitational point lens \citep{Jow:2020}. Likewise, the diffraction pattern of a binary gravitational lens strongly depends on the individual masses of the gravitating bodies \citep{Feldbrugge:2020}. The detection of fringes in the lensing of FRBs can tightly constrain the abundance of dark matter in the form of massive astrophysical compact halo objects (MACHOs) \citep{Katz:2020}. The observation of fringes in the lensing of gravitational waves can be used to study cosmological parameters \citep{Meena:2020, Tambalo:2022}. 



In the coming years, telescopes such as the Canadian Hydrogen Intensity Mapping Experiment (CHIME) \citep{CHIME:2018}, the Hydrogen Intensity and Real-time Analysis eXperiment (HIRAX) \citep{HIRAX:2016}, and the Square Kilometer Array (SKA) \citep{SKA:2015}, as well as next-generation follow-ups including the Canadian Hydrogen Observatory and Radio-transient Detector (CHORD) \citep{CHORD:2019} and the Packed Ultra-wideband Mapping Array (PUMA) \citep{PUMA:2019} will detect large numbers of coherent radio sources on the sky \citep{Petroff:2019}. Also, gravitationally lensed gravitational waves are expected to be observed in the near future, given the increasing number of detections and prospected increase in sensitivity of the VIRGO and LIGO detectors \citep{LIGO:2019}. Improving our understanding of wave effects in plasma and gravitational lensing will allow us to take advantage of these observations. 

In this paper, I provide a rigorous definition for multi-plane lensing in wave optics, and develop an efficient method to evaluate the corresponding oscillatory integral. I show that, unlike the geometric optics intensity, wave optics of multi-plane lensing is sensitive to the redshifts of the lens planes. In section \ref{sec:single}, I briefly discuss single-plane thin lenses in both geometric and wave optics, and describe the main features of Picard-Lefschetz theory and catastrophe theory following \cite{Feldbrugge:2018JCAP,Feldbrugge:2019}. In section \ref{sec:mutli}, I extend these techniques to the multi-plane thin lens system. In sections \ref{sec:plasma} and \ref{sec:gravitational}, I apply the new techniques to multi-plane plasma and gravitational lens systems.

\section{Single-plane lens} \label{sec:single}
A lens is a physical system in which radiation, for example, electromagnetic or gravitational, is deflected while traveling from the source to the observer. When the deflection takes place in a single region that is small compared to the distance between the source and observer, it can be modeled with a single-plane thin lens (see fig.\ \ref{fig:single}). 

The (relative) time delay in a single thin lens system, for a ray traveling from the source at an angular diameter distance $d_s$ and angular position $\bm{y}$ with respect to the reference point RP to the observer via the lens $L$ at an angular diameter distance $d_l$ and angular angular position $\bm{x}$ takes the form
\begin{align}
    T(\bm{x},\bm{y}) = \frac{1+z}{c} \frac{d_{l}d_{s}}{d_{sl}} \left(\frac{(\bm{x}-\bm{y})^2}{2} + \varphi(\bm{x})\right)\,,
\end{align}
with the redshift of the lens $z$, the angular diameter distance between the lens plane and the source plane $d_{sl}= d_s-d_l$, and the speed of light in vacuum $c$. The quadratic term captures the geometry of the light ray and the phase variation $\varphi$ describes the properties of the lens screen. For a gravitational lens with a surface mass density $\Sigma$, the gravitational phase variation $\varphi$ is given by 
\begin{align}
\varphi(\bm{x}) =-\frac{1}{\pi} \int  \kappa(\bm{x}') \ln\|\bm{x} - \bm{x}'\|\mathrm{d}\bm{x}'\,,
\end{align}
where
\begin{align}
\kappa(\bm{x}) = \frac{4 \pi G}{c^2} \frac{d_{l}d_{sl}}{d_{s}}\Sigma(d_{l}\bm{x})\,,
\end{align}
with Newton's constant $G$. For a plasma with an electron density $\Sigma_{e}$, the phase variation takes the form
\begin{align}
\varphi(\bm{x}) = - \frac{d_{l}d_{sl}}{d_{s}}\frac{\Sigma_{e}(\bm{x}) e^2}{m_e\, \epsilon_0\, \omega^2}\,,
\end{align}
with the electron mass $m_e$ and charge $e$, the vacuum permittivity $\epsilon_0$ and the frequency of the radiation $\omega$.


\subsection{Geometric optics}
In this section, we review the theory of geometric optics for the single-plane lens, and the role of caustics in the intensity patterns. This allows us to reflect on essential notation and key ideas, which we will later generalize to multi-plane lenses in wave optics. As we will see in the next section, it is at the caustics, where the geometric optics approximation fails, that wave effects become important.

Fermat's principle states that we observe images along rays for which the time delay is constant for all possible variations. The classical images are critical points of $T$ in the lens plane $L$ satisfying the equation
\begin{align}
    \nabla_{\bm{x}}T(\bm{x},\bm{y}) = \bm{x}-\bm{y} + \nabla \varphi(\bm{x}) = \bm{0}\,.\label{eq:Fermat}
\end{align}
This equation can have multiple solutions $\bar{\bm{x}}_i$ for fixed $\bm{y}$, each corresponding to a classical image reaching the observer from a distinct angle. Solving equation \eqref{eq:Fermat} for $\bm{y}$ for fixed $\bm{x}$ yields the Lagrangian map 
\begin{align}
    \bm{\xi}(\bm{x}) = \bm{x} + \nabla \varphi(\bm{x})\,.
\end{align}
A classical ray which passes the lens plane $L$ in $\bm{x}$ originated from the point $\bm{\xi}(\bm{x})$ on the source plane $S$. The gradient $\bm{\alpha} = \nabla\varphi$ is known as the deflection angle of the lens. The intensity in geometric optics can be expressed in terms of the Jacobian of the Lagrangian map $\mathcal{D} = \nabla \xi(\bm{x})$, 
\begin{align}
    I_{\text{geometric}}(\bm{y}) &\propto \sum_{\bm{x} \in \bm{\xi}^{-1}(\bm{y})} \frac{1}{|\det \nabla \bm{\xi}(\bm{x})|}\\
    &=
    \sum_{\bm{x} \in \bm{\xi}^{-1}(\bm{y})} \frac{1}{|\det (I + \mathcal{H}\varphi(\bm{x}))|}\\
    &=
    \sum_{\bm{x} \in \bm{\xi}^{-1}(\bm{y})} \frac{1}{|\det (\mathcal{H} T(\bm{x},\bm{y}))|}
\end{align} 
with the Hessian operator $\mathcal{H}$. The sum runs over the classical rays reaching the observer. 

\begin{figure}
    \centering
    \begin{subfigure}[b]{0.49\linewidth}
        \includegraphics[width =\textwidth]{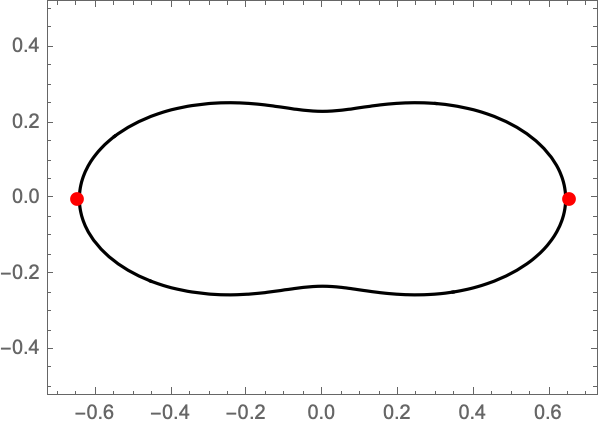}
        \caption{Critical curve $\mathcal{M}$}
    \end{subfigure}
    \begin{subfigure}[b]{0.49\linewidth}
        \includegraphics[width =\textwidth]{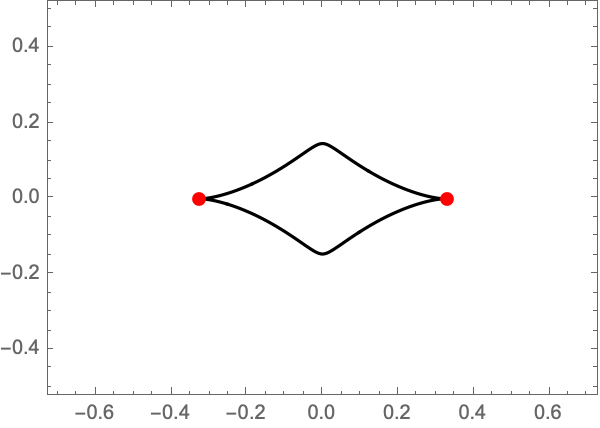}
        \caption{Caustic curve $\mathcal{C}$}
    \end{subfigure}
    \caption{The critical curve and caustic curve, with black the fold curve and red the cusp point, for the time delay $T(\bm{x},\bm{y})=\frac{(\bm{x}-\bm{y})^2}{2} +\frac{1}{2+2x_1^2+4 x_2^2}$.}\label{fig:ExampleCaustic}
\end{figure}

The intensity spikes in a caustic when the critical point is degenerate, \textit{i.e.}, the Hessian $\mathcal{H} T(\bm{x},\bm{y})$ is singular. Caustic are general features of lens systems occurring where the number of images changes. It is common to define the critical curve of a lens system as
\begin{align}
\mathcal{M} = \left\{ \bm{x} \in L \mid \det \nabla \bm{\xi} (\bm{x}) = 0\right\} .\label{eq:critical}
\end{align}
The intensity spikes at the caustics curve 
\begin{align}
    \mathcal{C} = \{\bm{\xi}(\bm{x})\,|\, \bm{x} \in \mathcal{M}\}\,.
\end{align}
For an example of the critical and caustic curve, consider the single-plane thin lens with the phase variation $\varphi(\bm{x})=1/(2+2x_1^2+4 x_2^2)$ where $\bm{x}=(x_1,x_2)$ (see fig.\ \ref{fig:ExampleCaustic}). The critical curve consists of a fold loop with two cusp points. The caustic curve consists of a loop with two points at which the fold loop is nondifferentiable. These points are the cusp points.

\begin{figure}
    \centering
    \begin{subfigure}[b]{0.32\linewidth}
        \includegraphics[width =\textwidth]{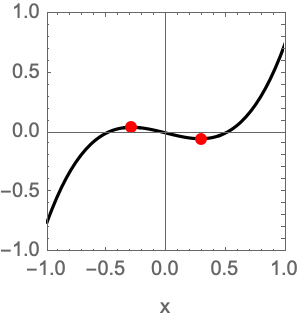}
        \caption{$y=-0.25$}
    \end{subfigure}
    \begin{subfigure}[b]{0.32\linewidth}
        \includegraphics[width =\textwidth]{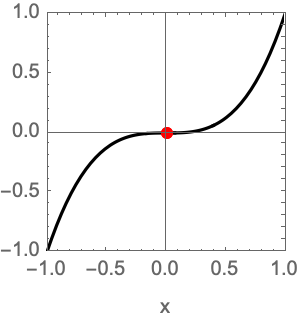}
        \caption{$y=0$}
    \end{subfigure}
    \begin{subfigure}[b]{0.32\linewidth}
        \includegraphics[width =\textwidth]{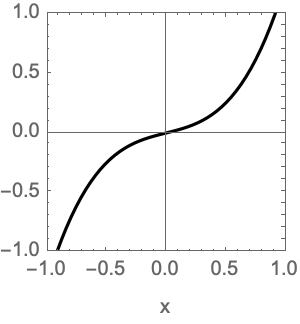}
        \caption{$y=0.25$}
    \end{subfigure}
    \caption{The unfolding of the critical point of $f=x^3 + x y$ with the critical points.}\label{fig:Catastrophe}
\end{figure}

Folds and cusps are examples of degenerate critical points. A generic function $f:\mathbb{R}^d \to \mathbb{R}$  only has non-degenerate critical points. That is, when a degenerate critical point occurs, it is not stable as a small perturbation to the function $f$ removes it. For example, the degenerate critical point $x=0$ of the function $f(x)=x^3$ is either removed or decomposes into a maximum and a minimum upon the addition of a small linear term $f(x) =x^3 + \epsilon x$ for small $\epsilon$ (see fig.\ \ref{fig:Catastrophe}). However, stable degenerate critical points do exist when we consider a family of functions $f_{\bm{y}}$ with the control parameter $\bm{y}$. Upon any small perturbation, the degenerate critical point persists if we allow for the control parameter to vary. For example, when considering the function $f(x)=x^3 + y x$, we find the stable degenerate critical point at $(x,y)=(0,0)$. A small deformation will move the critical point in the $xy$-plane. Depending on the dimensionality of the control parameter, a finite set of degenerate critical points fully classifies the stable possibilities. 
This classification is known as known as catastrophe theory, and is a remarkable extension of Morse's classification of non-degenerate critical points into maxima, minima, and saddle points \citep{Aronld:1973,Thom:1975,Berry:1980}. 

For lens systems, it is natural to associate the relative position of the source $\bm{y}$ with the control parameter of catastrophe theory. However, the control parameter can extend to any continuous deformation of the lens system. When considering a single control parameter, the stable degenerate critical points are known as fold caustics. These caustics correspond to a single vanishing eigenvalue, \textit{i.e.}, one of the eigenvalues of the deformation matrix $\mathcal{H}T$ vanishes $\lambda_i=0$, where the eigenvalues are defined by the eigenequation 
\begin{align}
    \mathcal{H}T\ \bm{v}_i = \lambda_i \bm{v}_i\,,    
\end{align}
with the eigenvector $\bm{v}_i$. The fold caustic marks the transition between regions in the control parameter space at which the number of images changes. When considering two control parameters, we can in addition to the fold caustic find stable cusp caustics. For the cusp caustic, the time delay function satisfies the additional condition
\begin{align}
    \bm{v}_i \cdot \nabla_{\bm{y}} \lambda_i = 0\,.
\end{align}
Not only the eigenvalue but also its derivative in the direction of the eigenvector field vanishes. For systems with more control parameters, the list of stable critical grows, including the swallowtail, butterfly and umbilic caustics. For a detailed exposition of these caustic conditions see \cite{Aronld:1973,Berry:1980,Blandford:1986,Hidding:2014,Feldbrugge:2018JCAP}.

\subsection{Wave optics}\label{sec:WaveOpticsSingle}
In wave optics, we do not study the evolution of classical rays. Instead, we express the intensity as the magnitude $|\Psi(\bm{y})|^2$ of the amplitude $\Psi(\bm{y})$ which follows from the Kirchhoff-Fresnel diffraction integral
\begin{align} \label{eq:Kirchhoff1d}
    \Psi(\bm{y}) = \mathcal{N} \int_{\mathbb{R}^d}e^{i\omega T(\bm{x},\bm{y})}\mathrm{d}\bm{x}\,,
\end{align} 
with the frequency of the radiation $\omega$. The integral is normalized with respect to the setup without a lens, $\varphi=0$, with the normalization constant
\begin{align}
    \mathcal{N} = \left(\frac{\omega}{2\pi i} \frac{1+z}{c} \frac{d_{l}d_{s}}{d_{sl}} \right)^{d/2}\,.
\end{align}
The Kirchhoff-Fresnel integral ranges over all possible light rays from the source to the observer and lets them interfere with each other. This is in fact a one-dimensional version of the Feynman path integral of quantum mechanics \citep{Feynman:1965}. 

The Kirchhoff-Fresnel integral is a conditionally convergent integral -- relying on a delicate cancelation of oscillations -- since the integrand is a pure phase and the integration domain has an infinite volume. In the following, we will use Picard-Lefschetz theory to unambiguously define this integral and evaluate it efficiently.

The Kirchhoff-Fresnel integral can be regulated in a variety of ways, which not necessarily yield the same answer. We here consider the class of smooth regulators $W_\sigma(\bm{x})$ which
\begin{enumerate}
\item[(i)] converge pointwise to $1$ as $\sigma \to \infty$, 
\item[(ii)] are analytic, \textit{i.e.}, do not have singularities in the complex $\bm{x}$-plane, and 
\item[(iii)] are integrable, \textit{i.e.}, $\int |W_\sigma(\bm{x})|\mathrm{d}\bm{x} < \infty$ for all $\sigma < \infty$. 
\end{enumerate}
The Gaussian regulator $W_\sigma(\bm{x}) = e^{-\bm{x}^2/(2\sigma^2)}$ and the $i\epsilon$ regulators are good candidates. Now we can define equation \eqref{eq:Kirchhoff1d} in the limit 
\begin{align}\label{eq:Kirchhoff1dreg}
    \Psi(\bm{y}) = \mathcal{N} \lim_{\sigma \to \infty} \int_{\mathbb{R}^d} e^{i \omega T(\bm{x},\bm{y})} W_\sigma(\bm{x})\mathrm{d}\bm{x}\,.
\end{align}
Note that the integral $\int e^{i \omega T(\bm{x},\bm{y})}W_\sigma(\bm{x})\mathrm{d}\bm{x}$ is absolutely convergent for $\sigma < \infty$, making the integral unambiguous. 

Cauchy's integral theorem states that the integral is invariant under a continuous deformation of the integration domain, as long as (i) we keep the endpoints fixed and (ii) we do not cross any singularities of the integrand. Picard-Lefschetz theory yields the optimal deformation removing all the oscillations from the integral. We can use this to evaluate equation \eqref{eq:Kirchhoff1dreg}.
First, we find the complex critical points $\bar{\bm{x}}_i$ of the analytic continuation of the time delay function $T(\bm{x},\bm{y})$ for a fixed $\bm{y}$. Since all critical points of an analytic function are saddle points by virtue of the Cauchy-Riemann equations, we will henceforth denote them as saddle points. Second, we find the corresponding steepest ascent $\mathcal{K}_i$ and steepest descent manifolds $\mathcal{J}_i$ corresponding to the height function $h(\bm{x}) = \text{Re}[i \omega T(\bm{x},\bm{y})]$. According to Picard-Lefschetz theory, we can write the amplitude
\begin{align}
    \Psi(\bm{y}) = \mathcal{N} \lim_{\sigma \to \infty}
    \sum_j n_j \int_{\mathcal{J}_j} e^{i \omega T(\bm{x},\bm{y})} W_\sigma(\bm{y})\mathrm{d}\bm{x}\,.
\end{align}
with the integer $n_i$ counting the intersection of the steepest ascent manifold $\mathcal{K}_i$ corresponding to the saddle point $\bar{\bm{x}}_i$ with the original integration domain $\mathbb{R}^d$. A saddle point of the time-delay function $\bar{\bm{x}}_i$ with a non-zero intersection $n_i$ is called a \textit{relevant saddle point}. The sum of the relevant steepest descent manifolds is known as the Lefschetz thimble. Real saddle points correspond to classical rays. Complex relevant saddle points play an important role in the vicinity of caustics \citep{Jow:2021}. As the steepest descent flow preserves the imaginary part of the exponent $H(\bm{x})=\text{Im}[i \omega T(\bm{x},\bm{y})]$ (an application of the Cauchy-Riemann equation), we have now rephrased the highly oscillatory conditionally convergent integral \eqref{eq:Kirchhoff1d} as the sum of a series of monotonically decreasing absolutely convergent integrals,
\begin{align}
    \Psi(\bm{y}) = \mathcal{N} \lim_{\sigma \to \infty}
    \sum_j n_j e^{i H[\bar{\bm{x}}_j]}\int_{\mathcal{J}_j} e^{h(\bm{x})} W_\sigma(\bm{y})\mathrm{d}\bm{x}\,.
\end{align}
Furthermore, as the integral along the steepest descent manifold generally converges absolutely, we can bring the limit $\sigma \to \infty$ inside the integral (using the dominated convergence theorem) yielding the regulated equation
\begin{align}\label{eq:Kirchhoff1dPL}
    \Psi(\bm{y}) =\mathcal{N}
    \sum_j n_j \int_{\mathcal{J}_j} e^{i \omega T(\bm{x},\bm{y})} \mathrm{d}\bm{x}\,.
\end{align}
Note that by virtue of Picard-Lefschetz theory, the smoothly regulated Kirchhoff-Fresnel integral is independent of the regulator and equivalent to an integral along the steepest descent manifolds of a set of relevant saddle points. See \cite{Feldbrugge:2022} for a more rigorous derivation of this result. After finding the steepest descent manifolds of the relevant saddle points, one can efficiently evaluate the integrals \citep{Feldbrugge:2019}. For details on the implementation of this Picard-Lefschetz integration method, its numerical implementation, and examples see \url{https://p-lpi.github.io}.

\begin{figure}
    \centering
    \begin{subfigure}[b]{0.49\linewidth}
        \includegraphics[width =\textwidth]{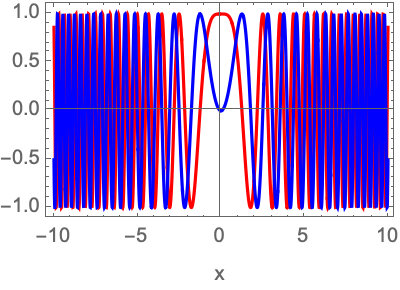}
        \caption{Original integrand $e^{ix^2}$}\label{fig:FresnelExampleL}
    \end{subfigure}
    \begin{subfigure}[b]{0.49\linewidth}
        \includegraphics[width =\textwidth]{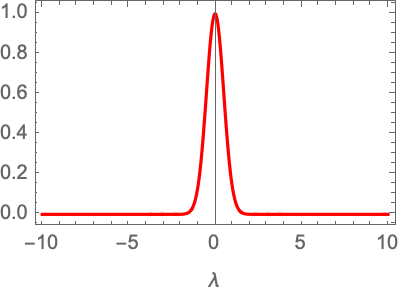}
        \caption{Deformed integrand $e^{-2\lambda^2}$}\label{fig:FresnelExampleR}
    \end{subfigure}
    \caption{The deformation of the Fresnel integral onto the steepest descent manifold. The real part in red and the imaginary part in blue.}\label{fig:FresnelExample}
\end{figure}

Let us observe how the procedure sketched above works in the example of the Fresnel integral $\int_{-\infty}^\infty e^{i x^2}\mathrm{d}x = \int_{-\infty}^\infty (\cos x^2 + i \sin x^2)\mathrm{d}x$. This integral is highly oscillatory and only conditionally convergent (see fig.\ \ref{fig:FresnelExampleL}). Adding a Gaussian regulator makes the integral absolutely convergent, \textit{i.e.}, $\int_{-\infty}^{\infty} e^{i x^2 - x^2/(2 \sigma^2)}\mathrm{d}x = \sqrt{\pi}/\sqrt{-i + 1/(2\sigma^{2})}$. Note that usually, such an exact solution does not exist. In the limit $\sigma \to \infty$, we obtain the result $(1+i)\sqrt{\pi/2}$. The exponent $i x^2$ in the integrand has a unique critical point at $x=0$, with the steepest descent manifold of the height function $h(x) = \text{Re}[ix^2]$ given by the rotated real line $\mathcal{J} = (1+i)\mathbb{R}$. Along the steepest descent manifold parameterized as $x = (1+i)\lambda$, the integrand transforms into the Gaussian $e^{i x^2} = e^{-2 \lambda^2}$ (see fig.\ \ref{fig:FresnelExampleR}). Integration along $\mathcal{J}$ yields the same result, \textit{i.e.}, $\int_\mathcal{J} e^{ix^2}\mathrm{d}x = (1+i) \int_{-\infty}^\infty e^{-2 \lambda^2}\mathrm{d}\lambda = (1+i) \sqrt{\pi/2}$. This illustrates that the outcome is independent of the regulator.

In the previous section, we noted that the geometric optics approximation fails at the caustics while transitioning between regions with a different number of classical images. Caustics are also important to wave optics itself. In definition \eqref{eq:Kirchhoff1dPL}, we observe that the Kirchhoff-Fresnel integral is governed by the relevant saddle points. As one approaches a caustic in the control parameter space $\bm{y}$, several non-degenerate critical points merge to form a degenerate critical point on the critical curve of the system (see eq.\ \eqref{eq:critical}). The fold caustic corresponds to the merger of two non-degenerate saddle points. The cusp caustic corresponds to the merger of three saddle points. After passing through the caustic, two of the saddle points transition into the complex plane forming a complex conjugate pair. Generally one of these complex saddle points remains relevant whereas the other saddle point is irrelevant to the Kirchhoff-Fresnel integral.

In the low-frequency domain, we can approximate the Kirchhoff-Fresnel integral with the perturbative expansion
\begin{align}
    \Psi(\bm{y}) &= \mathcal{N} \int_{\mathbb{R}^d}e^{i\omega T(\bm{x},\bm{y})}\mathrm{d}\bm{x}\\
    &= \mathcal{N} \int_{\mathbb{R}^d} e^{i \nu \frac{(\bm{x}-\bm{y})^2}{2}}
    \left(1+ i \nu \varphi(\bm{x}) + \mathcal{O}(\nu^2)\right)\mathrm{d}\bm{x}\\
    &= 1 +  i \mathcal{N}  \nu  \int_{\mathbb{R}^d} e^{i \nu  \frac{(\bm{x}-\bm{y})^2}{2}}
    \varphi(\bm{x}) \mathrm{d}\bm{x} + \mathcal{O}(\nu^2)\,,
\end{align}
with $\nu =  \frac{1+z}{c} \frac{d_{l}d_{s}}{d_{sl}} \omega$, which can be evaluated as a convolution with Fast Fourier Transforms. Geometric optics emerges as the WKB approximation of the Kirchhoff-Fresnel integral for the real critical points in the high-frequently limit. As $\nu$ increases, the integrand away from the critical points is exponentially suppressed. Locally approximating the exponent near the saddle points with a second-order Taylor approximation, the geometric optics approximation follows from the Gaussian integral. When including all relevant saddle points, we obtain the Eikonal or semi-classical approximation of Kirchhoff-Fresnel integral
\begin{align}
    \Psi_{\text{Eikonal}}(\bm{y})  = \sum_j  \frac{n_j e^{i \omega T(\bar{\bm{x}}_j,\bm{y})}}{\sqrt{\det (\mathcal{H}T(\bar{\bm{x}}_j,\bm{y}))}} \,,
\end{align}
and flux $I_\text{Eikonal} = |\Psi_\text{Eikonal}(\bm{y})|^2$. The Eikonal approximation includes the interference of rays in multi-image regions which is ignored in the geometric optics approximation. However, the Eikonal approximation still diverges near the caustics where the second-order Taylor expansion does not suffice. As we have seen, Picard-Lefschetz theory allows us to efficiently evaluate the lens integral without any approximations, enabling the study of lenses in the vicinity of caustics and bridging the perturbative and semi-classical regimes.

\section{Multi-plane lens} \label{sec:mutli}
So far, we have studied single-plane lens systems. In reality, lensing may be much more complicated than this, and include several lens planes. Consider radiation traveling through $n$ discrete lens planes $L_i$, $i=1,\dots,n$, at redshifts $z_i$ separating the observer from the source, such that $0 \leq z_1 \leq \dots \leq z_n$. Let $d_{i\,j}$ be the angular diameter distance between the lens spheres $L_i$ and $L_j$, and $\bm{x}_i$ the angular coordinate of the intersection of a light ray with the lens-plane $L_i$, with respect to the coordinate system centered on the observer and reference point RP (see Fig.\ \ref{fig:multi}). Let $\bm{y}=\bm{x}_{n+1}$ be the angular position at which the light ray intersects the source plane $S=L_{n+1}$. The lens system is again completely determined by the time-delay function (relative to the unperturbed trajectory)
\begin{align}
T(\bm{x}_1,\dots, \bm{x}_{n+1}) =& \sum_{j=1}^n T_j(\bm{x}_j,\bm{x}_{j+1}),
\label{eq:delay}
\end{align}
with the time-delay $T_j(\bm{x}_j,\bm{x}_{j+1})$ between planes $L_j$ and $L_{j+1}$ given by
\begin{align}\label{eq:timedelay}
    T_j(\bm{x}_j,\bm{x}_{j+1}) = 
    \frac{(1+ z_j)}{c} \frac{d_{0\,j}d_{0\, j+1}}{d_{j\, j+1}} \Big[ \frac{\left( \bm{x}_{j} - \bm{x}_{j+1}\right)^2}{2} + \beta_{j\,j+1}\varphi_j(\bm{x}_j)\Big]
\end{align}
\citep{Blandford:1986,Schneider:1992}. The first term is the Pythagorean contribution, corresponding to the variations in the path length. The second is the time delay due to a gravitational- or plasma-induced phase $\varphi_j$, received while passing through the $j$-th lens plane, weighted by the geometric coefficient $\beta_{j\,j+1}$ with 
\begin{align}
\beta_{i\,j} = \frac{d_{i\,j} d_{0\,n+1}}{d_{0\,j}d_{i\,n+1}}.
\end{align}
Given a surface mass density $\Sigma_j$, the gravitational phase variation $\varphi_j(\bm{x}_j)$ is given by 
\begin{align}
\varphi_j(\bm{x}_j) =-\frac{1}{\pi} \int \mathrm{d}\bm{x}' \kappa_j(\bm{x}') \ln\|\bm{x}_j - \bm{x}'\|\,,
\end{align}
where
\begin{align}
\kappa_j(\bm{x}_j) = \frac{4 \pi G}{c^2} \frac{d_{0\, j}d_{j\,n+1}}{d_{0\,n+1}}\Sigma_j(d_{0\,j}\bm{x}_j)\,,
\end{align}
with Newton's constant $G$. For a plasma with an electron density $\Sigma_{j,e}$ on $L_j$, the phase variation is
\begin{align}
\varphi_j(\bm{x}_j) = - \frac{d_{0\,j}d_{j\,n+1}}{d_{0\,n+1}}\frac{\Sigma_{j,e}(\bm{x}_j) e^2}{m_e\, \epsilon_0\, \omega^2},
\end{align}
with the electron mass $m_e$ and charge $e$, the vacuum permittivity $\epsilon_0$ and the frequency of the radiation $\omega$.

\subsection{Geometric optics}
The geometric approximation for multi-plane lenses elegantly extends the approximation for single-plane lenses. Fermat's principle applied to the time-delay function \eqref{eq:delay}, yields the conditions
\begin{align}
\bm{x}_{j+1} - \bm{x}_{j}
=&
-\frac{1+z_{j-1}}{c} \frac{d_{0\,{j-1}}d_{j\, j+1}}{d_{0\,j+1}d_{j-1\,j}} \left( \bm{x}_{j-1} - \bm{x}_{j}\right)\nonumber\\
&+\beta_{j\,j+1}\bm{\alpha}_j\left(\bm{x}_j\right),\quad j=1,\dots,n
\end{align}
with $\bm{x}_0=\bm{0}$, and the deflection angle $\bm{\alpha}_j \equiv \nabla \varphi_j$. This iterative equation is solved by
\begin{align}
\bm{x}_j = \bm{x}_1 + \sum_{i=1}^{j-1} \beta_{i\,j} \bm{\alpha}_i(\bm{x}_i),\ j=2,\dots, n+1\,,
\end{align}
and induces the Lagrangian map $\bm{\xi}:L_1 \to S$.  that expresses the intersection of the ray with the source plane in terms of the angle on the sky,
\begin{align}
\bm{\xi}(\bm{x}_1) = \bm{x}_1 + \sum_{i=1}^{n} \beta_{i\,n+1} \bm{\alpha}_i(\bm{x}_i).
\end{align}
Note that geometric optics is completely determined by the geometric coefficients $\beta_{i\,j}$ and the deflection angles $\bm{\alpha}_i$. It does not depend on the redshifts $z_i$ of the lens planes.

In geometric optics, the deformation tensor $\mathcal{D} = \nabla_{\bm{x}_1} \bm{\xi}$ determines the lensing pattern. Whereas the deformation tensor of a single-plane lens ($n=1$) is symmetric, including shear and expansion terms, the deformation tensor of the multi-plane lens is generically not symmetric. In this case, the deformation tensor additionally includes a rotation. The intensity of the image is given by
\begin{align}
I_\text{geometric}(\bm{y}) = \sum_{\bm{x}_1 \in \bm{\xi}^{-1}(\bm{\mu})} \frac{1}{|\det \nabla_{\bm{x}_1} \bm{\xi} (\bm{x}_1)|}
\end{align}
which spikes on the caustic curve 
\begin{align}
    \mathcal{C} = \{ \bm{\xi}(\bm{x}_1)\,|\, \bm{x}_1 \in \mathcal{M}\}\,,
\end{align}
with the critical curve
\begin{align}
\mathcal{M} = \left\{ \bm{x}_1 \in L_1 \mid \det \nabla_{\bm{x}_1} \bm{\xi} (\bm{x}_1) = 0\right\}.
\end{align}
The one- and two-dimensional cases, respectively, generally lead to the occurrence of fold caustics, and fold and cusp caustics.

\subsection{Wave optics} \label{sec:wo}
In this section, we introduce a new technique to evaluate multi-plane lensing in wave optics. In wave optics, the lens pattern is again determined by the time-delay function, but this time through the higher-dimensional Kirchhoff-Fresnel integral,
\begin{align}
\Psi(\bm{y}) = \mathcal{N} \int_{(\mathbb{R}^{d})^n} e^{i \omega T(\bm{x}_1,\dots,\bm{x}_n,\bm{y})}\mathrm{d}\bm{x}_1\dots\mathrm{d}\bm{x}_n, \label{eq:KirchhoffFresnelMulti}
\end{align}
evaluated over piecewise linear paths between the source, the lens planes, and the observer, with the frequency $\omega$. The integral is normalized with the normalization constant
\begin{align}
    \mathcal{N} = \prod_{j=1}^n \mathcal{N}_j \,,
\end{align}
with
\begin{align}
    \mathcal{N}_j = \left(\frac{\omega}{2\pi i}\frac{1+z_j}{c} \frac{d_{0\,j}d_{0\, j+1}}{ d_{j\, j+1}}\right)^{d/2},
\end{align}
ensuring that that the intensity $I(\bm{y}) = |\Psi(\bm{y})|^2$ is unity in the absence of a lens, \textit{i.e.}, $I(\bm{y})=1$ when $\varphi_j = 0$ for all $j$. This multi-plane Kirchhoff-Fresnel integral was recently derived using path integral methods and analytically studied for the case of two aligned gravitational point lenses \citep{Yamamoto:2017}. In this paper, we solve the problem for more general configurations leveraging more advanced integration methods.

In theory, we could apply Picard-Lefschetz theory directly to the highly oscillatory $nd$-dimensional integral $\Psi$. However, direct evaluation is unpractical due to its high dimensionality. Recently, we have seen the development of Hamiltonian Monte Carlo methods for high-dimensional integration using Picard-Lefschetz theory \citep{Cristoforetti:2012,Cristoforetti:2013,Tanizaki:2014, Alexandru:2015}. These Monte Carlo methods are very powerful for a small set of high-dimensional integrals. Here we instead provide a scheme to efficiently evaluate a large family of the Kirchhoff-Fresnel integrals as a function of the position of the source $\bm{y}$. We do this by making clever use of the special structure of the time delay function $T$.\\

\noindent To evaluate the Kirchhoff-Fresnel integral \eqref{eq:KirchhoffFresnelMulti}, we convert the $nd$-dimensional integral into $n$ $d$-dimensional integrals
\begin{align} \label{eq:nynke}
\Psi_{i+1}(\bm{x}_{i+1})=\mathcal{N}_i\int_{\mathbb{R}^d} \Psi_{i}(\bm{x}_{i})e^{i \omega T_i(\bm{x}_i,\bm{x}_{i+1})}\mathrm{d}\bm{x}_i\,,
\end{align}
for $i=1,\dots, n$, with the initial wave function $\Psi_1(\bm{x}_1)=1$, the final amplitude $\Psi(\bm{y}) =\Psi_{n+1}(\bm{y})$. Using Cauchy's integral theorem, we deform the integration domain into the complex plane. For the first integral, $\Psi_{2}(\bm{x}_{2})$, we deform the integration domain $\mathbb{R}^d$ to the Lefschetz thimble $\mathcal{J}_1 \subset \mathbb{C}^d$ corresponding to a set of relevant saddle points of the exponent $i \omega T_1(\bm{x}_1,\bm{x}_2)$. As we saw in section \ref{sec:WaveOpticsSingle}, along the Lefschetz thimble, the integral is absolutely convergent and relatively easy to evaluate,
\begin{align}
\Psi_{2}(\bm{x}_{2})=\mathcal{N}_1\int_{\mathcal{J}_1} e^{i \omega T_1(\bm{x}_1,\bm{x}_{2})}\mathrm{d}\bm{x}_1\,.
\end{align}
Assume that the phase variation $\varphi_1$ consists of a set of localized features. Away from the caustics, the dominant relevant saddle point is given by the unperturbed ray $\bm{x}_1 = \bm{x}_2$. We thus find an approximation of the diffraction integral in the single image regions
\begin{align}
\Psi_{2}(\bm{x}_{2})
&\approx e^{i \omega T_1(\bm{x}_2,\bm{x}_2)}\\
&= e^{i \omega
\frac{1+z_1}{c} \frac{d_{0\,1}d_{0\, 2}}{d_{1\, 2}} \beta_{1\,2}\varphi_1\left(\bm{x}_2\right)
}\,.
\end{align}$ $\\
\indent Next, express the amplitude $\Psi_2$ as the sum of this single image approximation and a term containing the details of the interference pattern in the multi-image regions,
\begin{align}
\Psi_2(\bm{x}_2) = e^{i\omega T_1(\bm{x}_2,\bm{x}_2)} + \delta \Psi_2(\bm{x}_2)\,.
\end{align}
Note that $\delta \Psi_2(\bm{x}_2)$ decays exponentially in the single image regions away from the caustics. Using this expansion, we write the second integral as
\begin{align}
\Psi_{3}(\bm{x}_{3})= \mathcal{N}_2 \bigg[&\int_{\mathbb{R}^d} e^{i \omega \left(T_1(\bm{x}_2,\bm{x}_2) + T_2(\bm{x}_2,\bm{x}_{3})\right)}\mathrm{d}\bm{x}_2\nonumber\\
&+ \int_{\mathbb{R}^d} \delta\Psi_2(\bm{x}_2) e^{i \omega T_2(\bm{x}_2,\bm{x}_{3})}\mathrm{d}\bm{x}_2 \bigg]\,.
\end{align}
The first integral can be efficiently evaluated with Picard-Lefschetz theory, deforming the integration domain to the Lefschetz thimble $\mathcal{J}_2\subset \mathbb{C}^d$ of the second integration variable $\bm{x}_2$. We evaluate the second integral with conventional integration methods. Since the fluctuation $\delta\Psi_2$ decays exponentially away from the caustics, it suppresses the highly oscillatory character of the integral and makes it absolutely convergent. In particular, in this paper, we use Fast Fourier Transforms to evaluate the convolution with the Gaussian kernel $\exp\left[i \omega \frac{1+z}{c} \frac{d_{0\ j} d_{0\ j+1}}{d_{j\ j+1}} \frac{\bm{x}^2}{2}\right]$. 

Again writing the amplitude $\Psi_3$ as a perturbation with respect to the trivial saddle points $\bm{x}_2=\bm{x}_3$, \textit{i.e.}, $\Psi_3(\bm{x}_3) = e^{i\omega(T_1(\bm{x}_3,\bm{x}_3)+T_2(\bm{x}_3,\bm{x}_3))} + \delta \Psi_3(\bm{x}_3)$, we can write the subsequent integral as single-plane Kirchhoff-Fresnel integral plus an absolutely convergent correction. 
We iterate this procedure till we reach the desired amplitude $\Psi(\bm{y}) = \Psi_{n+1}(\bm{y})$.

Note that the Kirchhoff-Fresnel integral exhibits two interesting limits. Firstly, for small separations between the lens planes $d_{j\, j+1}$, the geometric term dominates over the phase variation $\varphi_j$. In the limit where $d_{j\, j+1}$ approaches $0$, the geometric term approaches a Dirac delta function in the integral, collapsing the lens planes $L_j$ and $L_{j+1}$ to a new plane with a phase variation $\varphi_j + \varphi_{j+1}$. Secondly, when the phase variations $\varphi_j$ and $\varphi_{j+1}$ of two lens planes $L_j$ and $L_{j+1}$ have disjoint support, the lens integral factorizes into the product of two Fresnel-Kirchhoff integrals.

Unlike in the geometric approximation, the diffraction pattern of the multi-plane lens system in wave optics is sensitive to the redshifts of the lens planes $z_j$. The fringes in wave optics thus, in principle, can be used to probe the redshifts of the lens planes in a novel way.

\begin{figure*}
    \centering
    \begin{subfigure}[b]{0.49\linewidth}
        \includegraphics[width =\textwidth]{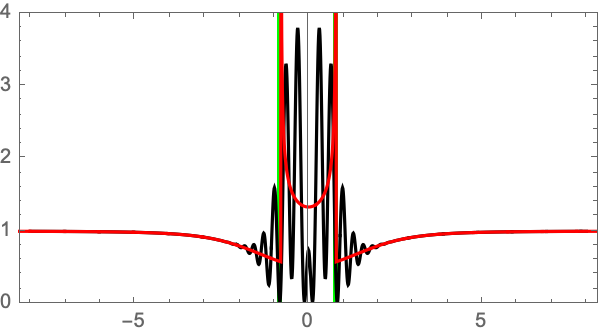}
        \caption{$|\Psi_2(x_2)|^2$}\label{fig:Example1L}
    \end{subfigure}
    \begin{subfigure}[b]{0.49\linewidth}
        \includegraphics[width =\textwidth]{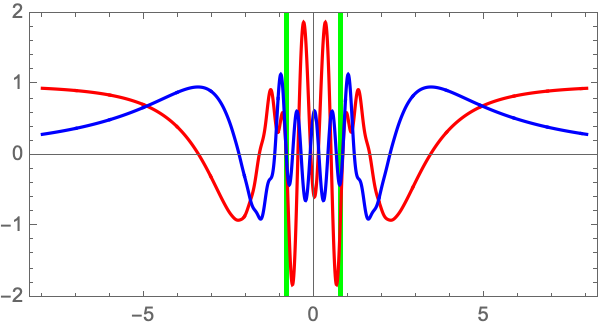}
        \caption{$\Psi_2(x_2)$}\label{fig:Example1R}
    \end{subfigure}\\
    \begin{subfigure}[b]{0.49\linewidth}
        \includegraphics[width =\textwidth]{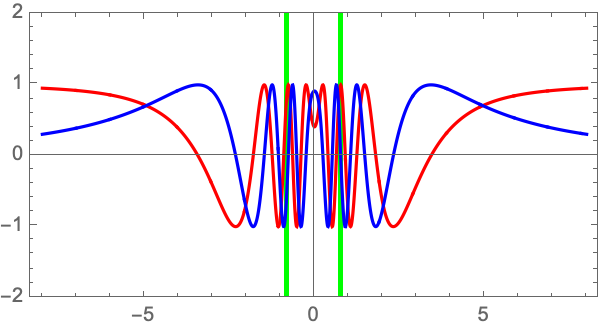}
        \caption{$\exp[i \omega T_1(x_2,x_2)]$}\label{fig:Example2L}
    \end{subfigure}
    \begin{subfigure}[b]{0.49\linewidth}
        \includegraphics[width =\textwidth]{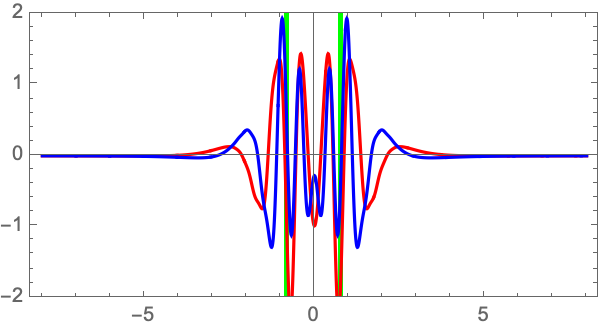}
        \caption{$\delta \Psi_2(x_2)$}\label{fig:Example2R}
    \end{subfigure}\\
    \begin{subfigure}[b]{0.49\linewidth}
        \includegraphics[width =\textwidth]{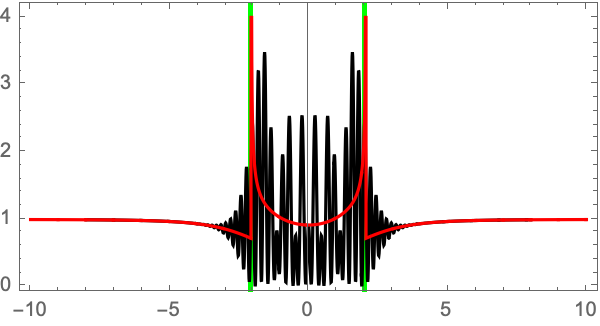}
        \caption{$|I_1(y)|^2$}\label{fig:Example3L}
    \end{subfigure}
    \begin{subfigure}[b]{0.49\linewidth}
        \includegraphics[width =\textwidth]{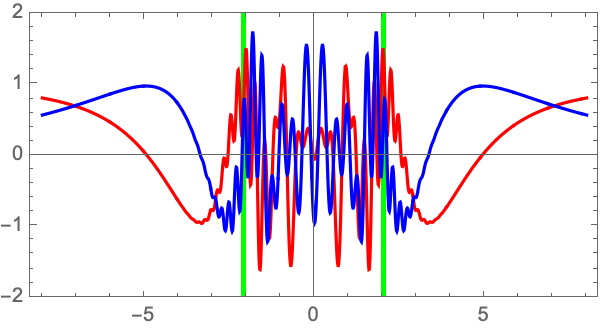}
        \caption{$I_1(y)$}\label{fig:Example3R}
    \end{subfigure}\\
    \begin{subfigure}[b]{0.49\linewidth}
        \includegraphics[width =\textwidth]{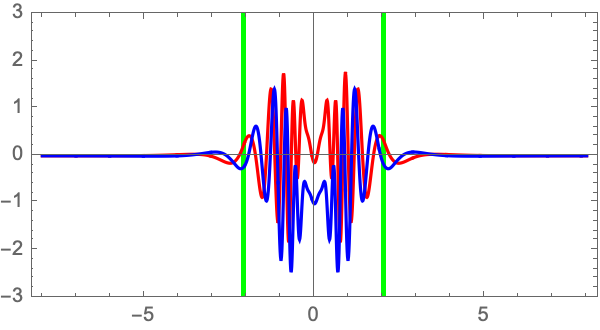}
        \caption{$I_2(x_2)$}\label{fig:Example4L}
    \end{subfigure}
    \begin{subfigure}[b]{0.49\linewidth}
        \includegraphics[width  =\textwidth]{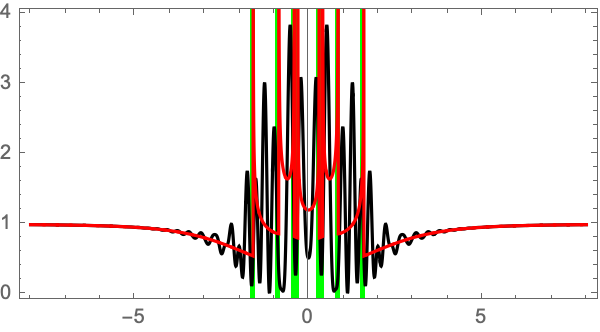}
        \caption{$|\Psi(y)|^2$}\label{fig:Example4R}
    \end{subfigure}
    \caption{The intermediate steps in the evaluation of the one-dimensional rational double plane lens system for $\omega=10$. The modulus squared of the amplitude (in black) with the geometric optics approximation (in red). The real and imaginary parts of the amplitude (in red and blue). The green vertical lines mark the caustics.}\label{fig:Example}
\end{figure*}

\subsection{Example: double rational lens}
\begin{figure*}
    \centering
    \begin{subfigure}[b]{0.33\linewidth}
        \includegraphics[width =\textwidth]{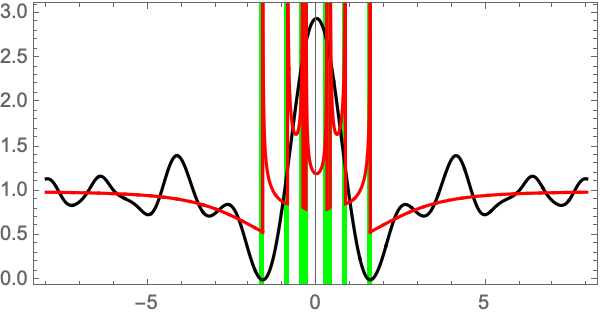}
        \caption{$\omega=1$}
    \end{subfigure}
    \begin{subfigure}[b]{0.33\linewidth}
        \includegraphics[width =\textwidth]{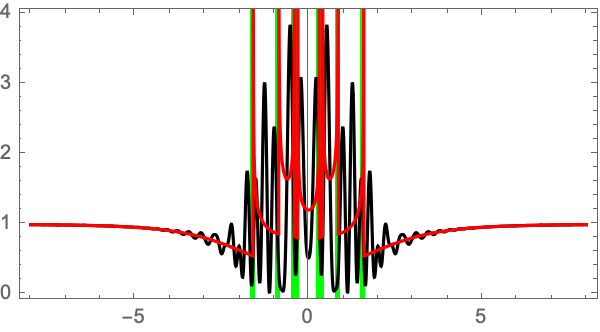}
        \caption{$\omega=10$}
    \end{subfigure}
    \begin{subfigure}[b]{0.33\linewidth}
        \includegraphics[width =\textwidth]{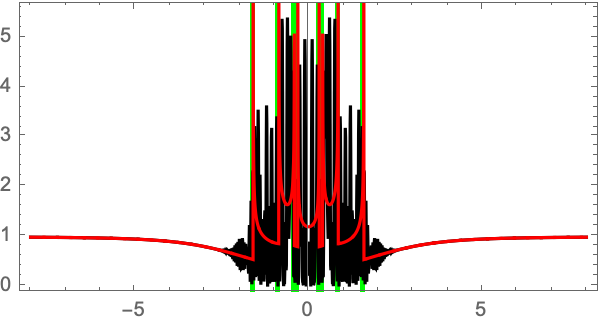}
        \caption{$\omega=50$}
    \end{subfigure}
    \caption{The flux of the double-plane system for three frequencies. The modulus squared of the amplitude (in black) with the geometric optics approximation (in red). The green vertical lines mark the caustics.}\label{fig:Example_freq}
\end{figure*}
I demonstrate these techniques explicitly for a one-dimensional example. Consider the double-plane lens system, with the time delay functions
\begin{align}
    T_1(x_1,x_2) &= \frac{(x_1-x_2)^2}{2} + \frac{\alpha_1}{1+x_1^2}\,, \\
    T_2(x_2,y) &= \frac{(x_2-y)^2}{2} + \frac{\alpha_2}{1+x_2^2}\,,
\end{align}
corresponding to two co-linear rational thin lenses. Fermat's principle yields the conditions
\begin{align}
    0&=x_1-x_2 - \frac{2 x_1 \alpha_1}{(1+x_1^2)^2}\,,\\
    0&=2 x_2 - x_1 - y-\frac{2 x_2 \alpha_2}{(1+x_2^2)^2}\,.
\end{align}
Solving this set of equations for $x_2$ and $y$ yields the Lagrangian map,
\begin{align}
    \xi_2(x_1) &= x_1 - \frac{2 x_1 \alpha_1}{(1+x_1^2)^2}\,,\\
    \xi(x_1) &= 2 \xi_2(x_1)-x_1 - \frac{2 \xi_2(x_1) \alpha_2}{(1+\xi_2(x_1)^2)^2}\,.
\end{align}
The flux in geometric optics takes the form 
\begin{align}
    I_\text{geometric}(y) &\propto \sum_{x_1 \in \xi^{-1}(y)} \frac{1}{|\det \partial_{x_1} \xi(x_1)|}\,,
\end{align}
with a series of caustics corresponding to the zero crossings of the Jacobian $\partial_{x_1} \xi(x_1)$.
In wave optics, the Kirchhoff-Fresnel integral is again fully determined by the time delay function
\begin{align}
    \Psi(y) = \frac{\omega}{2\pi i} \iint e^{i \omega \left(T_1(x_1, x_2) + T_2(x_2, y)\right)}\mathrm{d}x_1 \mathrm{d}x_2\,.
\end{align}
To evaluate this integral, first we express the two-dimensional integral as two one-dimensional integrals 
\begin{align}
    \Psi_2(x_2) &= \sqrt{\frac{\omega}{2\pi i}} \int e^{i \omega T_1(x_1,x_2)}\mathrm{d}x_1\,,\\
    \Psi(y) &= \sqrt{\frac{\omega}{2\pi i}} \int \Psi_2(x_2)\, e^{i\omega T_2(x_2,y)}\mathrm{d}x_2\,.
\end{align}
Next, we evaluate the first integral $\Psi_2$ using Picard-Lefschetz theory. See fig.\ \ref{fig:Example1L} and \ref{fig:Example1R} for the flux with the corresponding geometric optics approximation and the real and imaginary parts of the amplitude. The amplitude $\Psi_2$ is well approximated in the single image regions by the  single-image approximation $\Psi_2(x_2) \approx e^{i \omega T_1(x_2,x_2)} = e^{i \omega \alpha_1/(1+x_2^2)}$ (see fig.\ \ref{fig:Example2L}). The residue decays rapidly in the single-image regions (see fig.\ \ref{fig:Example2R}). Substituting $\Psi_2$ expressed as $e^{i \omega T_1(x_2,x_2)} + \delta \Psi_2(x_2)$ into the second integral yields,
\begin{align}
    \Psi(y) = \sqrt{\frac{\omega}{2\pi i}}\ \bigg[&\int e^{i \omega(T_1(x_2,x_2) + T_2(x_2,y))} \mathrm{d}x_2\nonumber\\
    & +
    \int \delta \Psi_2(x_2) e^{i \omega T_2(x_2,y)} \mathrm{d}x_2 \bigg]\,.
\end{align} 
The first integral 
\begin{align}
    I_1 = \sqrt{\frac{\omega}{2\pi i}}\int \exp \left[i \omega\left(\frac{(x_2-y)^2}{2} + \frac{\alpha_1+\alpha_2}{1+x_2^2}\right)\right] \mathrm{d}x_2
\end{align}
can be evaluated with Picard-Lefschetz theory. See fig.\ \ref{fig:Example3L} and \ref{fig:Example3R} for the flux with the corresponding geometric optics approximation and the real and imaginary parts of the amplitude. As the second integral 
\begin{align}
    \sqrt{\frac{\omega}{2\pi i}}\int \delta\Psi_2(x_2)\exp\left[i\omega \left(\frac{\alpha_2}{1+x_2^2} + \frac{(x_2-y)^2}{2}\right)\right]\mathrm{d}x_2\,.
\end{align}
is exponentially suppressed for large $x_2$, the oscillations remain contained and the integral is numerically tractable with conventional integration techniques. For clarity, we plot the identity $I_2(x_2) = \delta\Psi_2(x_2)\exp(i\omega \alpha_2/(1+x_2^2))$ in fig.\ \ref{fig:Example4L}. We can evaluate this second integral as a convolution of $I$ with a Gaussian kernel with two Fast Fourier transforms,
\begin{align}
    \int I_2(x_2) e^{i\omega (x_2-y)^2/2}\mathrm{d}x_2 = \text{FFT}^{-1}\left[ FFT[I_2] e^{-i k^2/(2 \omega)}\right]\,.
\end{align}
When we add these two contributions we find the solution to the Kirchhoff-Fresnel integral. See fig.\ \ref{fig:Example4R} for the resulting flux in relation to the corresponding geometric optics approximation. Note that while the intermediate calculations only included single- and triple-image regions, the full multi-plane lens consists of single-, triple-, as well as quintuple-image regions. The interference pattern is richer and only emerges when taking all the relevant contributions into account.

As the frequency increases, the wave optics calculation moves from the perturbative to the semi-classical regime approaching the geometric optics approximation (see fig.\ \ref{fig:Example_freq}). For the frequency $\omega=1$, we are in the perturbative regime. The intensity does not resolve the caustics of the system. For the frequently $\omega=50$, we are approaching the geometric optics regime. Away from the caustics, the amplitude is well approximated by the Eikonal method.

\section{Application: Plasma lensing}\label{sec:plasma}

\begin{figure*}
    \begin{subfigure}[b]{0.33\textwidth}
        \includegraphics[width =\textwidth]{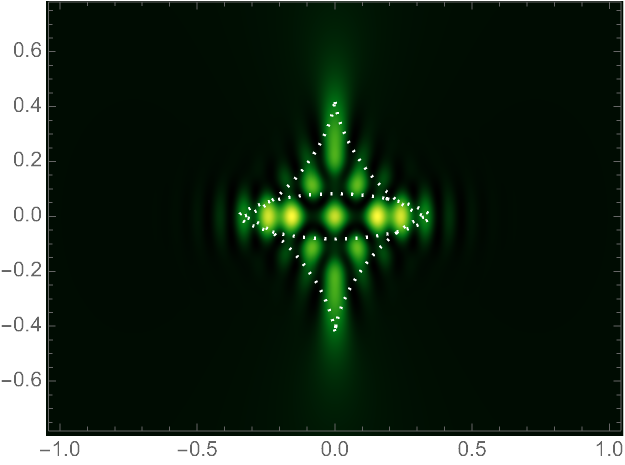}
    \caption{$I(\bm{x}_2)=\big|\Psi_2(\bm{x}_2)\big|^2$}\label{fig:Plasma1}
    \end{subfigure}
    \begin{subfigure}[b]{0.33\textwidth}
        \includegraphics[width =\textwidth]{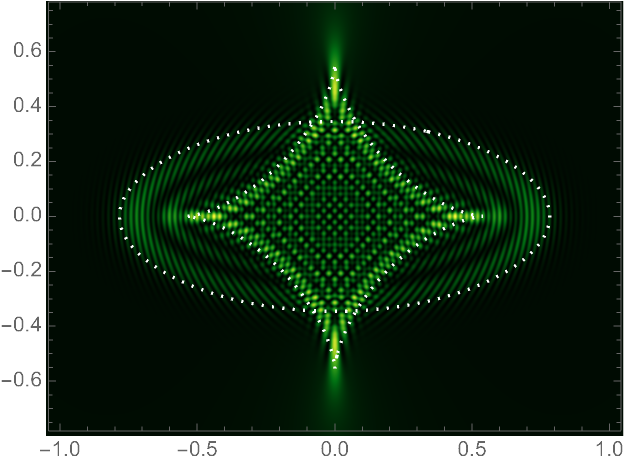}
        \caption{$I(\bm{x}_3)=\big|\int_{\mathbb{R}^d} e^{i \omega \left(T_1 + T_2\right)}\mathrm{d}\bm{x}_2\big|^2$}\label{fig:Plasma2}
    \end{subfigure}
    \begin{subfigure}[b]{0.33\textwidth}
        \includegraphics[width =\textwidth]{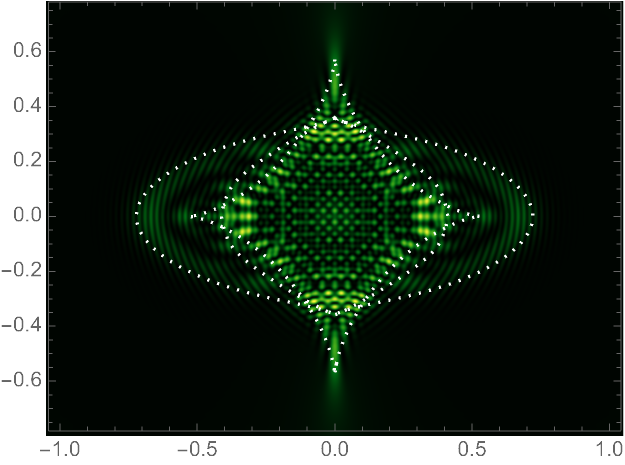}
        \caption{$I(\bm{x}_3)=\big|\Psi_3(\bm{x}_3)\big|^2$}\label{fig:Plasma3}
    \end{subfigure}
    \caption{Example of a plasma lens with the frequency $\omega = 50$. The white dotted line is the caustic curve. \emph{Left:} intensity pattern on the first lens-plane $L_1$. \emph{Center:} single-plane approximation on the second lens-plane $L_2$. \emph{Right:} double lens system. }\label{fig:Plasma}
\end{figure*}

For a further demonstration of the method introduced in section \ref{sec:wo}, consider a two-dimensional double plasma lens system ($d=n=2$) with two rational phase variations 
\begin{align}\label{eq:rationallens}
    \varphi_i(x,y) = \frac{1}{1+ 2x^2 + y^2}\,,
\end{align}
for $i=1,2$. Let the lens planes be equally spaced between the source and the observe, $d_{0\,1}=d_{1\,2}=d_{2\,3} = 1/3$. For simplicity, we assume the lenses to be relatively close to the observer, with redshifts $z_1=z_2=0$. Phase variations \eqref{eq:rationallens} constitute a technically convenient model as the analytic continuation of the time-delay function for such a lens has a finite number of saddle points in the complex $\mathbb{C}^2$ plane. The resulting flux is qualitatively similar to that of the Gaussian lens model 
\begin{align}
    \varphi_i(x,y) = e^{-x^2-2 y^2},
\end{align}
but the analytic continuation of the latter would include an infinite set of complex saddle points.

The double-plane rational plasma model consists of a single-image region, multiple triple-image regions, and a single quintuple-image region (see fig.\ \ref{fig:Plasma3}). The first lens leads to an interference pattern that is still close to the perturbative regime (see fig.\ \ref{fig:Plasma1}). The double-plane lens in the single-image approximation for the first lens yields a more intricate result (see fig.\ \ref{fig:Plasma2}). When combining the single -image contribution with the multi-image contribution, the interference pattern changes qualitatively as the caustic structure changes.

\begin{figure*}
    \begin{subfigure}[b]{0.44\linewidth}
        \includegraphics[width =\textwidth]{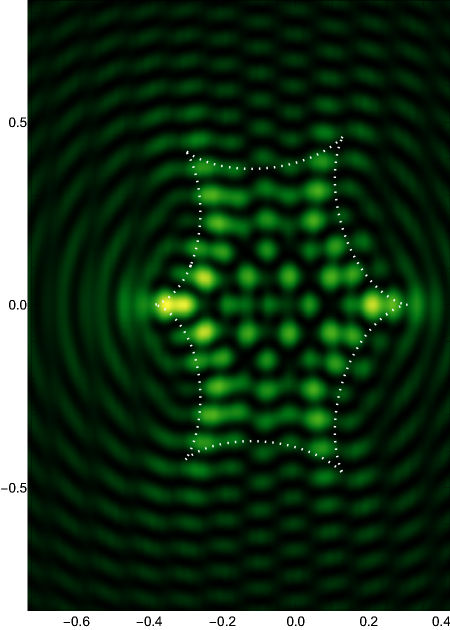}
        \caption{Binary lens with a small radial separation.}
    \end{subfigure}
    \begin{subfigure}[b]{0.44\linewidth}
        \includegraphics[width =\textwidth]{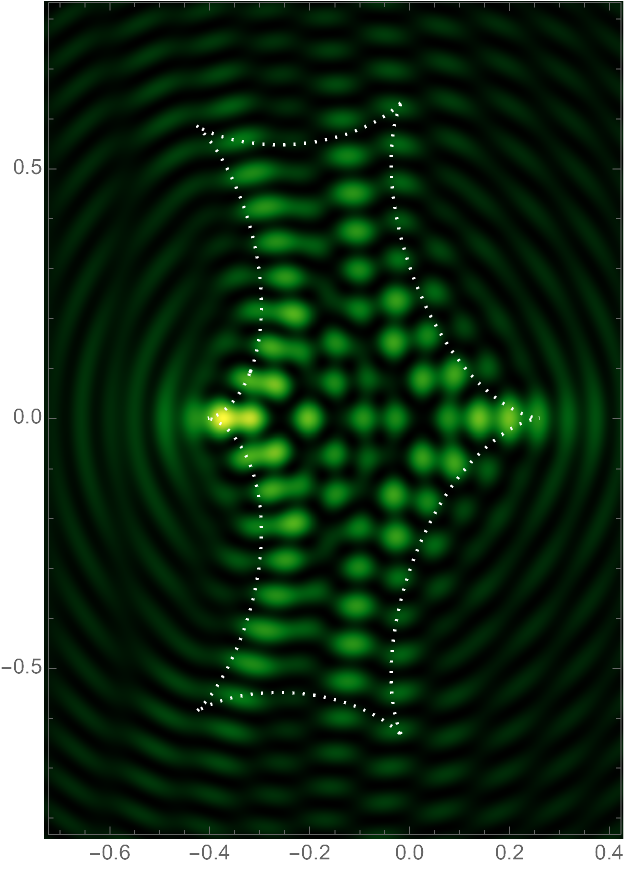}
        \caption{Binary lens in a single lens plane.}
    \end{subfigure}
    \caption{
        Binary system separated by an Einstein angle with the mass weightings $g_1=\frac{1}{3},g_2=\frac{2}{3}$ and the frequency $\omega = 50$. The white dotted line is the caustic curve. \textit{Left:} the masses have a small radial separation. \textit{Right:} the masses are in a single plane.
        }\label{fig:double}
\end{figure*}

\begin{figure*}
    \begin{subfigure}[b]{0.32\textwidth}
        \includegraphics[width =\textwidth]{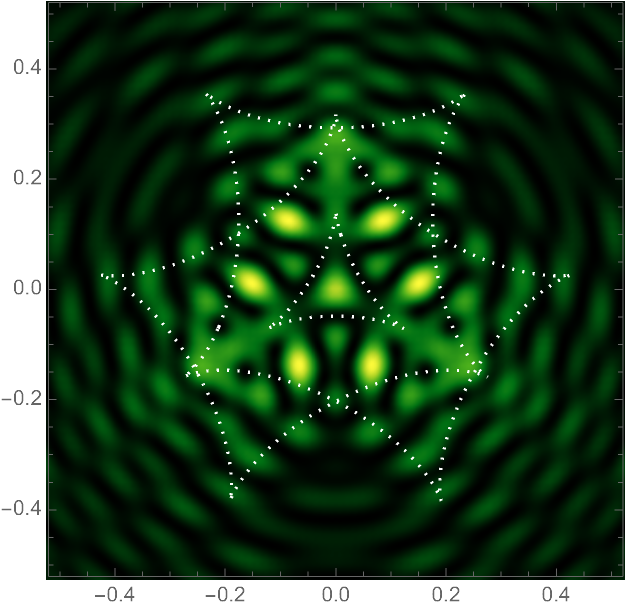}
        \caption{$\omega=50$}
    \end{subfigure}
    \begin{subfigure}[b]{0.32\textwidth}
        \includegraphics[width =\textwidth]{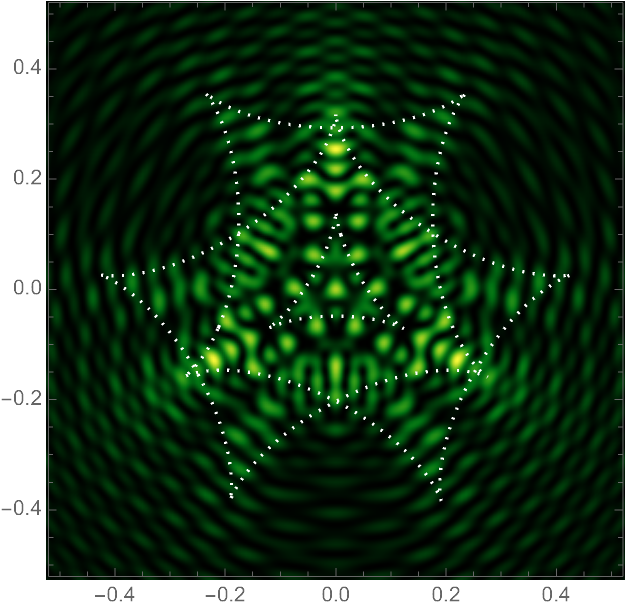}
        \caption{$\omega=100$}
    \end{subfigure}
    \begin{subfigure}[b]{0.32\textwidth}
        \includegraphics[width =\textwidth]{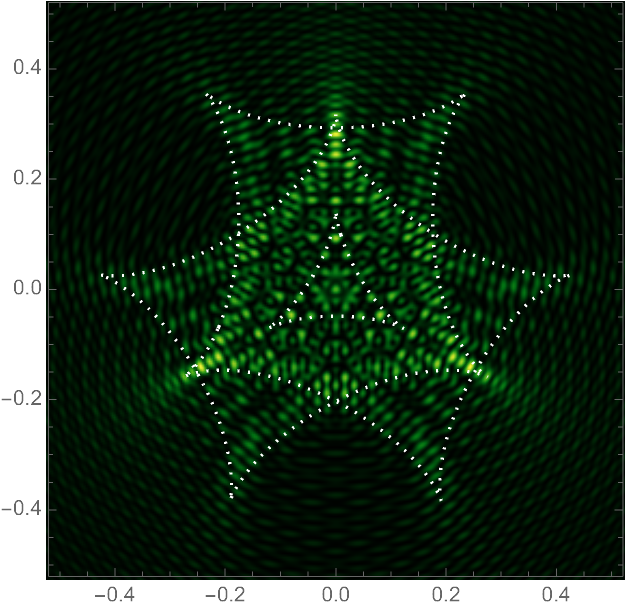}
        \caption{$\omega=200$}
    \end{subfigure}
    \caption{Triple point mass lens interference pattern for the frequencies $\omega = 50,100$, and $200$. The white dotted line is the caustic curve.}\label{fig:triple}
\end{figure*}

\section{Application: Gravitational lensing} \label{sec:gravitational}
The proposed iterative procedure works efficiently for smooth lenses. Gravitational lenses for which the phase variation is singular are more subtle. A massive body, where we assume that the mass is concentrated in a point, located at the angular position $\bm{r}$ induces a phase variation which goes as the logarithm of the angular distance $-\ln \| \bm{x} - \bm{r}\|$. The norm and the branch cut of the analytic continuation of the logarithm make the Picard-Lefschetz analysis of such a system more intricate. Nonetheless, the methods developed in the previous sections still apply when combined with the exact solution of the single-body gravitational lens and elliptic coordinates for the binary gravitational lens system \citep{Feldbrugge:2020}.

Consider a setup with $n$ point masses with mass $M_i$ in the lens planes $L_i$ for $i=1,\dots,n$. Let the $i$-th lens be located at the angular positions $\bm{r}_i$ in the lens planes  $L_i$. The $i$-th point mass induces the phase variation
\begin{align}
\varphi_i(\bm{x}_i - \bm{r}_i) = -g_i \ln\|\bm{x}_i - \bm{r}_i\|,
\end{align}
with the weighting $g_i = \frac{4 G M_i}{c^2} \frac{d_{i\,n+1}}{d_{0\,i}d_{0\,n+1}}$. The Kirchhoff-Fresnel integral \eqref{eq:KirchhoffFresnelMulti}, is again written as $n$ $2$-dimensional integrals. The first integral $\Psi_2$ has a closed form solution in terms of hypergeometric functions \citep{Nakamura:1999}. Writing 
\begin{align}
    \Psi_2(\bm{x}_2) = e^{-i \omega g_1 \ln\|\bm{x}_2-\bm{r}_1\|} + \delta \Psi_2(\bm{x}_2),
\end{align}
where the term $\delta \Psi_2$ decays exponentially away from the source, we express the second integral $\Psi_3$ as the sum of two terms
\begin{align}
    \Psi_3(\bm{x}_3) &= \mathcal{N}_2 \int_{\mathbb{R}^2} e^{i \omega \left(T_1(\bm{x}_2,\bm{x}_2)+T_2(\bm{x}_2,\bm{x}_3)\right)} \mathrm{d}\bm{x}_2\\
    &\phantom{=} +  
    \mathcal{N}_2 \int_{\mathbb{R}^2} \delta \Psi_2(\bm{x}_2)\ e^{i \omega T_2(\bm{x}_2,\bm{x}_3)} \mathrm{d}\bm{x}_2\,.
\end{align}
The first integral can be evaluated as a binary single-plane lens problem with elliptic coordinates (for the details see \citep{Feldbrugge:2020}). The second integral converges absolutely and can be efficiently evaluated with Fast Fourier Transforms. See fig.\ \ref{fig:double} for an illustration of two binary gravitational lens systems with the separation of the gravitating bodies $\|\bm{r}_1-\bm{r}_2\|=1$ in units of the Einstein angle of the binary system. In the left panel, the lens planes are separated by a small angular diameter distance ($d_{0\,1}=0.475$, $d_{1\,2} = 0.05$, $d_{2\,3}=0.475$). In the right panel, the two gravitating bodies lay in a single lens plane. Note that both the caustic cuves and the diffraction pattern significantly depend on this radial displacement. For an exploration of the caustics of gravitational binary lens systems in the geometric optics limit, see \cite{Erdl:1993}.\\

\noindent Let's now add a third gravitational lens to the setup. If we write the amplitude $\Psi_3$ in terms of a single-image approximation and a multi-image correction
\begin{align}
\Psi_3(\bm{x}_3) = e^{i\omega\left( T_1(\bm{x}_3,\bm{x}_3)+ T_2(\bm{x}_3,\bm{x}_3)\right)} + \delta \Psi_3(\bm{x}_3)\,,
\end{align}
we cannot directly use Picard-Lefschetz theory to evaluate the integral over the single-image approximation in $\Psi_4$. For the binary gravitational lens problem, we relied on the elliptic coordinates \citep{Feldbrugge:2020} to accommodate the singularities of the phase variation. Unfortunately, this does not straightforwardly generalize to problems with three gravitating bodies. The triple body problem includes three singularities located at $\bm{r}_1,\bm{r}_2,\bm{r}_3$ whereas an ellipse has only two foci which can be used to accommodate them. For this reason, we instead use the effective phase variation
\begin{align}
    - \left(\sum_{i=1}^2 g_i \right)\ln\|\bm{x} - \bm{r}_m\|\,,
\end{align}
with the weighted mean $\bm{r}_m = \frac{\sum_{i=1}^2 g_i \bm{r}_i}{ \sum_{i=1}^2 g_i}$. This is a good approximation of the sum $\sum_{i=1}^2 T_i$ away from the singularities. Using this effective phase, we express the amplitude $\Psi_3$ as
\begin{align}
\Psi_3(\bm{x}_3) = e^{- i \omega \sum_{i=1}^2 g_i \ln\|\bm{x}_3 - \bm{r}_m\|} +\delta\Psi_3(\bm{x}_3).
\end{align}
Using this expansion, the integral $\Psi_4$ splits into an integral which is analogous to the binary gravitational lens problem -- which can be solved with elliptic coordinates -- and an absolutely convergent correction corresponding to the interference pattern in the multi-image region. Note that it is merely a convenient representation of the calculation to add a new point mass lens. The result is still exact. We can add a fourth lens to the problem by iterating this procedure. 

Using this technique, we now can, for the first time, evaluate the triple single-plane gravitational lens problem. Let the three masses form an equilateral triangle $\bm{r}_1 = (1/2,-1/(2\sqrt{3}))$, $\bm{r}_2=(-1/2,-1/(2\sqrt{3}))$, $\bm{r}_3=(0,1/\sqrt{3})$, with sides equal to unity in terms of the Einstein angle, centered on $\frac{1}{3}\left(\bm{r}_1+\bm{r}_2+\bm{r}_3\right) = \bm{0}$. Furthermore, let the lens planes be located at the angular diameter distances $d_{0\,1}=1/2, d_{1\, 2} = d_{2\, 3}=0, d_{3\,4}=1/2$, $g_1=g_2=g_3=1/3$. Using elliptic coordinates for the first two gravitational lenses and adding the third lens on a separate plane in the limit $d_{2\, 3}\to 0$, we obtain an intricate interference pattern with single-, triple-, quintuple- and septuple-image regions (see fig.\ \ref{fig:triple}). Note that this diffraction pattern consists of a single four-image region, six six-image regions, and a single eight-image region enclosing a ten-image region. For the frequency $\omega=50$, not all the caustics are resolved in wave optics. As the frequency increases, the flux in the multi-image regions becomes more oscillatory and the caustic pattern emerges.

\section{Conclusion}
In this paper, I define and evaluate the diffraction patterns of multi-plane lensing in wave optics using a combination of Picard-Lefschetz theory and Fast Fourier Transforms. I express the integral as an iterated integral and write the intermediate integrals as the sum of a simple (unnormalized) Eikonal term and a multi-image term with compact support containing the interesting caustic and interference effects. This procedure is robust and converges in polynomial time. The quantum interference is significantly richer than geometric optics near the caustics where the geometric optics approximation breaks down. The diffraction pattern contains more features and is, unlike the geometric analysis, directly sensitive to the redshifts of the lens planes. In future work, I will perform a detailed investigation of the dependence of the fringes on the redshifts of the lens planes. This method furthers the study of the interplay of classical catastrophe theory and interference of coherent radio sources.

\section*{Data Availability}
No new data were generated or analyzed in support of this research.

\section*{Acknowledgements}
The author thanks Ue-Li Pen for the helpful discussions, and Neil Turok and Nynke Niezink for their comments on the manuscript. JF is supported in part by the Higgs Fellowship.

\bibliographystyle{mnras_sjf}
\bibliography{biblio} 



\appendix


\bsp	
\label{lastpage}
\end{document}